\documentclass[prl,showpacs,floatfix,twocolumn]{revtex4}

\usepackage{graphicx}
\usepackage{dcolumn}
\usepackage{bm}
\usepackage{natbib}

\begin{document}

\title{\boldmath Superconducting coherence lengths of hole-doped cuprates obtained from electron-boson spectral density functions \unboldmath}

\author{Jungseek Hwang}
\email{Electronic address: jungseek@skku.edu}
\affiliation{Department of Physics, Sungkyunkwan University, Suwon, Gyeonggi-do 16419, Republic of Korea}

\date{\today}

\begin{abstract}

Electron–boson spectral density functions (EBSDFs) can be obtained from measured spectra using various spectroscopic techniques, including optical spectroscopy. EBSDFs, known as glue functions, have a magnetic origin. Here, we investigated EBSDFs obtained from the measured optical spectra of hole-doped cuprates with wide doping levels, from underdoped to overdoped cuprates. The average frequency of an EBSDF provides the timescale for the spin fluctuations to form Cooper pairs. This timescale is directly associated with retarded interactions between electrons. Using this timescale and Fermi velocity, a reasonable superconducting coherence length, which reflects the size of the Cooper pair, can be extracted. The obtained coherence lengths were consistent with those measured via other experimental techniques. Therefore, the formation of Cooper pairs in cuprates can be explained by spin fluctuations, the timescales of which appear in EBSDFs. Consequently, EBSDFs provide crucial information on the timescale of the microscopic mechanism of Cooper pair formation.

\end{abstract}


\maketitle

Since the discovery of copper-oxide superconductors (or cuprates)\cite{bednorz:1986}, intensive experimental and theoretical studies have been performed on these intriguing material systems. The primary objectives of such studies included determining the microscopic pairing mechanism of the Cooper pairs in cuprate systems. Based on the generic phase of cuprates, antiferromagnetic spin fluctuations are expected to play an important role in the formation of Cooper pairs in such systems. One important result might be the electron–boson spectral density function (EBSDF) obtained from spectra measured using various spectroscopic techniques, including optical spectroscopy\cite{carbotte:1999,schachinger:2000,hwang:2006,zasadzinski:2006,lee:2006,valla:2007,hwang:2007,hwang:2007ab,heumen:2009,hwang:2011,carbotte:2011,zasadzinski:2011}.
This function, in principle, describes the interaction of two electrons through the exchange of force-mediating bosons; thus, the EBSDF has also been called the glue function\cite{heumen:2009}.The experimentally obtained EBSDFs are known to have a magnetic origin, that is, antiferromagnetic spin fluctuations\cite{carbotte:1999,schachinger:2000,johnson:2001,zasadzinski:2001,hwang:2004,norman:2004,hwang:2006,hwang:2007,dahm:2009}. This quantity can be denoted as $I^2B(\omega)$, where $I$ is the couple constant between the electron and the boson and $B(\omega)$ is the boson spectrum. Therefore, this glue function may carry crucial information on the formation of Cooper pairs. EBSDFs can be extracted from measured reflectance spectra using a well-developed procedure\cite{schachinger:2000,hwang:2006,hwang:2007,heumen:2009,heumen:2009a,hwang:2015a}. From the obtained EBSDFs, the coupling constants and the superconducting transition temperatures ($T_c$) can be determined using a generalized McMillan formula\cite{hwang:2008c,heumen:2009,hwang:2011}. Because the maximum possible Tc obtained from the experimental EBSDFs using the McMillan formula are higher than the actual superconducting transition temperatures, EBSDFs may be sufficient for realizing superconductivity. However, the microscopic mechanism of Cooper pair formation in cuprates has not yet been elucidated. Therefore, determining the microscopic paring mechanism has been an important and debated issue in contemporary condensed matter physics.

In this study, we demonstrate that the superconducting coherence length can be obtained from the experimental EBSDFs of cuprates at various doping levels (from underdoped to overdoped). The experimental EBSDFs were obtained from the measured optical spectra of Bi$_2$Sr$_2$CaCu$_2$O$_{8+\delta}$ (Bi-2212) and YBa$_2$Cu$_3$O$_{6.50}$ (Y123-orthoII). First, we obtained the average frequency of the EBSDF, which provides the timescale for the retarded interaction between two electrons in a Cooper pair. The coherence length is directly related to the size of the Cooper pair, and is determined by the Fermi velocity and timescale obtained from the experimental EBSDF. We also considered both the two-dimensional transport character and the anisotropic ($d$-wave) superconducting gap to obtain the $ab$-plane coherence length. The resulting coherence lengths at various doping levels are reasonable, compared with the coherence lengths measured using other experimental techniques.

Based on this study, we propose a microscopic pairing mechanism that is conceptually similar to that of conventional Bardeen–Cooper–Schrieffer (BCS) superconductors. However, there are some differences between the two superconductor types. The force-mediating boson is associated with spin fluctuations above the background (antiferromagnetic) spin in cuprates, whereas the boson is associated with the charge fluctuations above the background (lattice ion) charge in BCS superconductors. Spin fluctuations in cuprates play the same role as charge fluctuations in BCS superconductors. Therefore, magnetic retarded interaction  is the force required to form Cooper pairs in cuprates. The coherence lengths of cuprates are much shorter (usually two orders of magnitude) than those of BCS superconductors. This can be understood by considering that spin fluctuations have much shorter timescales compared with those of charge fluctuations, and the Fermi velocities of cuprates are slower than those of BCS superconductors.

\begin{figure}[!htbp]
  \vspace*{-1.0 cm}%
  \centerline{\includegraphics[width=3.5 in]{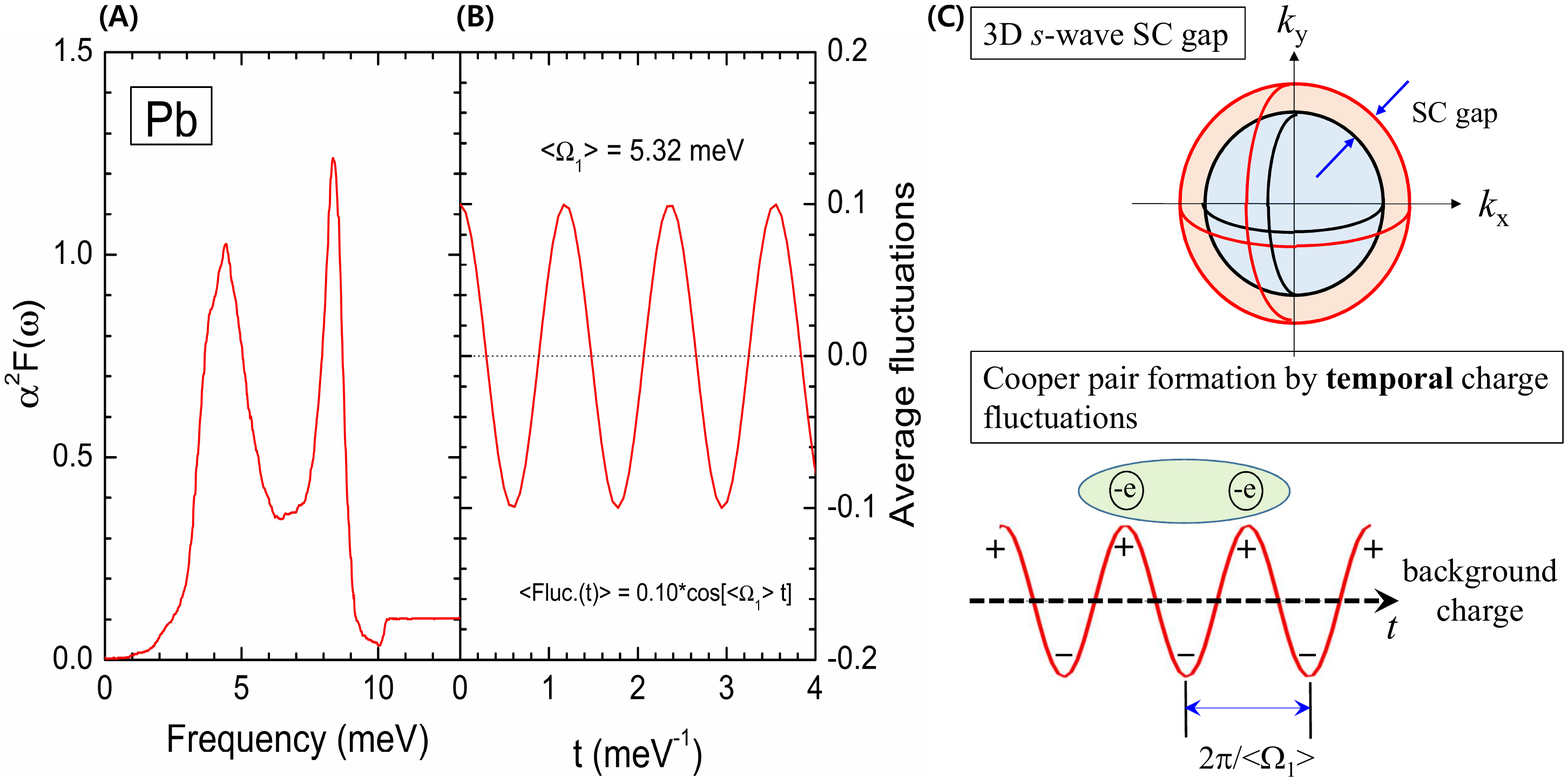}}%
  \vspace*{-1.0 cm}%
\caption{(Color online) (Color online) A conventional Bardeen–Cooper–Schrieffer (BCS) superconductor, Pb. (A) Electron–phonon spectral density function ($\alpha^2 F(\omega)$) of Pb. (B) Average charge fluctuations on top of the regular ion background charge. (C) Schematic diagrams for the $s$-wave superconducting (SC) gap and the Copper pair formation by the average charge fluctuations. Here $+$ and $-$ represent the positive and negative excess charge above the background charge, respectively.}
\label{fig1}
\end{figure}

Let us first consider a conventional BCS superconductor, Pb, which has been thoroughly studied. The electron–phonon spectral density function (EPSDF), $\alpha^2F(\omega)$, of Pb was obtained using theoretical calculations and experimental spectroscopic techniques, including optical spectroscopy\cite{mcmillan:1965,farnworth:1974,farnworth:1976,tomlinson:1976}. The obtained EPSDF of Pb is shown in Fig. 1(A). The EPSDF of Pb was used in this study to estimate the superconducting coherence length. First, the average frequency ($\langle\Omega_1\rangle$) of the EPSDF was calculated, defined as $\langle\Omega_1\rangle \equiv (2/\lambda)\int_{0}^{\omega_c} [\alpha^2F(\omega)/\omega] \omega d\omega$, where $\lambda$ is the coupling constant given by $\lambda \equiv 2\int_{0}^{\omega_c} [\alpha^2F(\omega)/\omega] d\omega$, with a cutoff frequency of $\omega_c$. The inverse of the average frequency ($2\pi/\langle\Omega_1\rangle$) is the average period of charge fluctuations in the EPSDF. The average charge fluctuation ($Fluc_{charge}(t)$) is described as $Fluc_{charge} = 0.10 \cos(\langle\Omega_1\rangle t)$. Note that the amplitude of the fluctuation is arbitrary. The fluctuating charge with the average frequency is shown in Fig. 1(B). The period of the fluctuating charge can be the timescale for the retarded interaction between two electrons in a Cooper pair. In Fig. 1(C) a schematic of the s-wave superconducting gap and Cooper pair formation by the {\it temporal} charge fluctuations is presented. As electrons have negative charge, two electrons can be paired at the maxima of the positive charge of the temporal fluctuations through the retarded interaction. The Coulomb repulsion between electrons in a pair can be overcome through this retarded attractive electric interaction. The coherence length ($\xi$) can be obtained using the timescale ($2\pi/\langle\Omega_1\rangle$) and the Fermi velocity ($v_{F}$). Because conventional BCS superconductors have three-dimensional (isotropic) $s$-wave gaps, an electron can be paired with any other electron on a sphere with the electron at the center. Consequently, spatial degeneracy increases the pairing probability by a factor of 4$\pi$, which is the full solid angle in three-dimensional space. This results in a coherence length that is shorter by the same factor. Therefore, the coherence length can be written as $\xi = v_F [2\pi/\langle\Omega_1\rangle](1/4\pi) = v_F/(2\langle\Omega_1\rangle)$. The Fermi velocity of Pb is $\sim$1.83$\times$10$^{6}$ m/s\cite{lykken:1971,ashcroft}. The estimated coherence length is 113 nm, which is consistent with the reported value of 96 nm\cite{gasparovic:1970}. Therefore, this approach is adequate for estimating the coherence length of a conventional superconductor, Pb. Further, this picture shows that Cooper pairs can be formed by a combination of the electron velocity at the Fermi surface and the timescale of the charge fluctuations. This explains why the Cooper pairs have much larger sizes compared with the spacings between electrons in the material system. Using this approach, we demonstrated that one of two characteristic length scales (coherence length and London penetration depth) for superconductivity can be determined from the EPSDF. Note that the EPSDF clearly carries information on the timescale of the retarded interaction.

\begin{figure}[!htbp]
  \vspace*{-0.8 cm}%
  \centerline{\includegraphics[width=3.7 in]{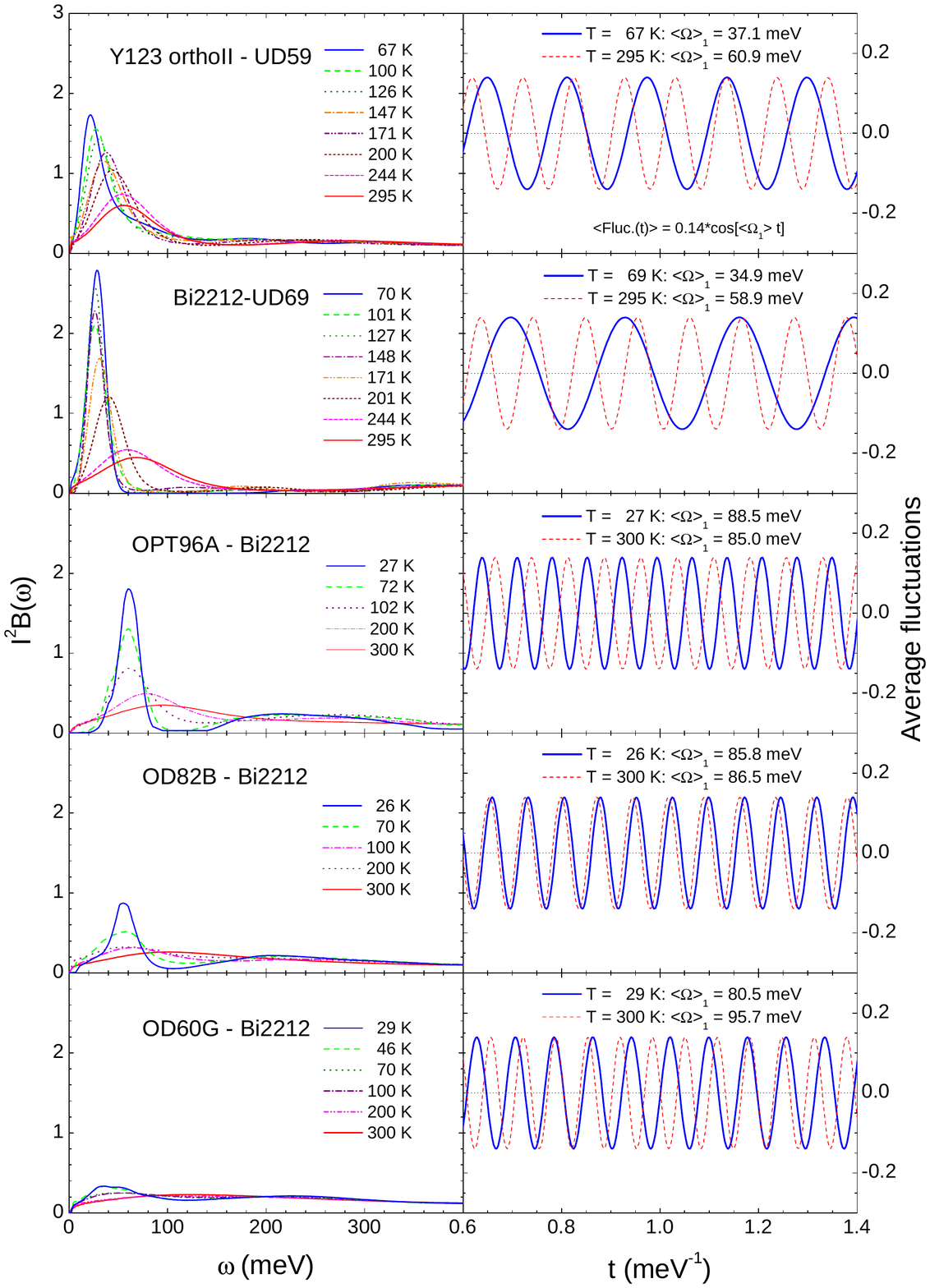}}%
  \vspace*{-1.0 cm}%
\caption{(Color online) Electron–boson spectral density functions (EBSDFs) of cuprates and their average frequencies. (Left column) EBSDFs ($I^2B(\omega)$) of cuprates at various doping levels and temperatures. (Right column) Average fluctuating spins of cuprates with the average frequencies at two (lowest and room) temperatures.}
\label{fig2}
\end{figure}

\begin{figure}[!htbp]
  \vspace*{-0.8 cm}%
  \centerline{\includegraphics[width=4.3 in]{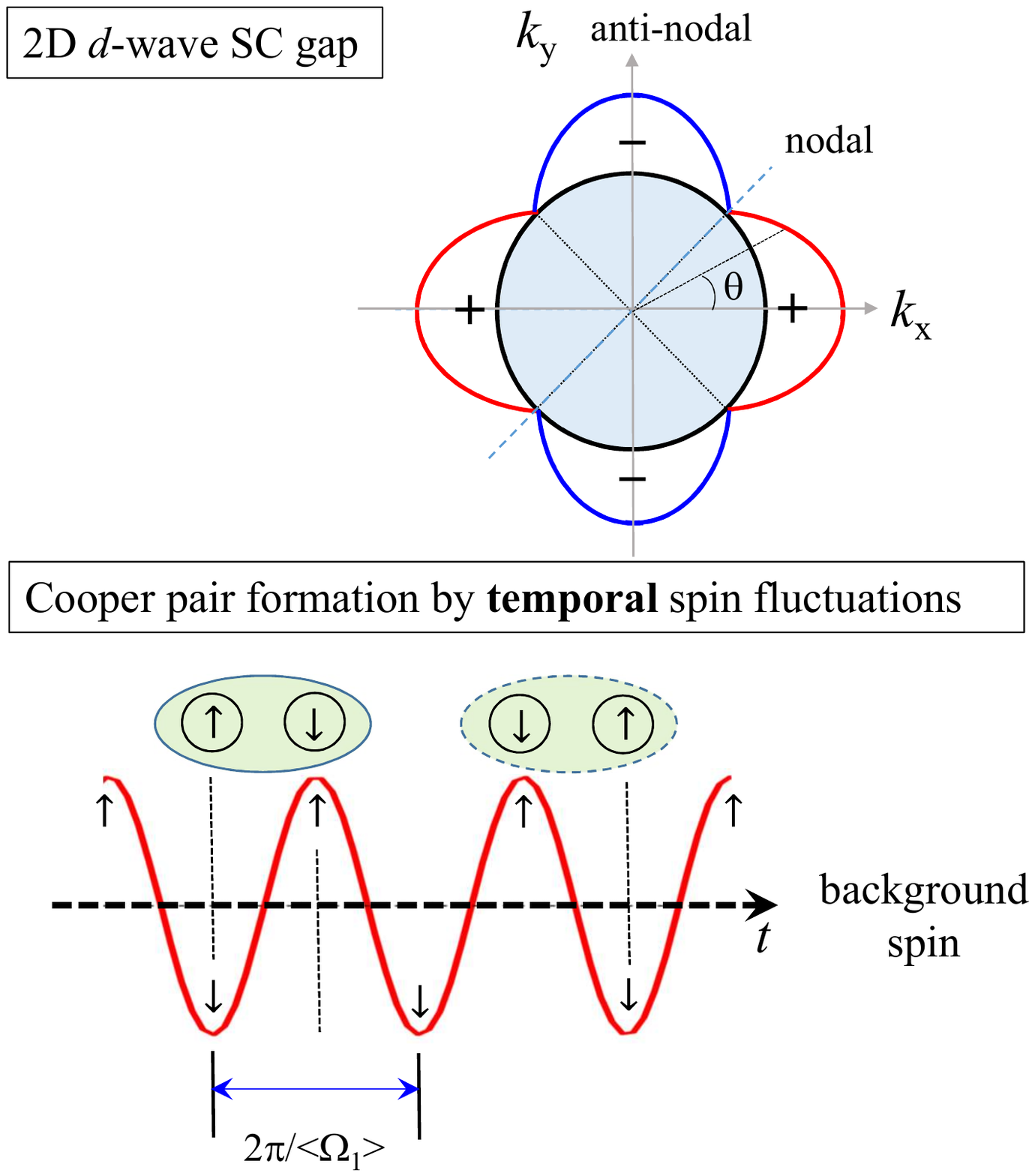}}%
  \vspace*{- 1.1 cm}%
\caption{(Color online) High-transition-temperature superconductors, cuprates. (Upper) Schematic of the d-wave superconducting (SC) gap. (Lower) Cooper pair formation by the average spin fluctuations. Here, $\uparrow$ and $\downarrow$ indicate the up  and down excess spins above the background spin, respectively.}
\label{fig3}
\end{figure}

Next, we applied the previously described approach to cuprate systems. We used the EBSDFs of cuprates obtained from optically measured spectra\cite{hwang:2008,hwang:2016a}. The EBSDFs of YBa$_2$Cu$_3$O$_{6.50}$ orthoII (Y123 orthoII) with $T_c =$ 59 K at various temperatures are obtained by using a maximum entropy method\cite{schachinger:2006,hwang:2016a} (also see Supplementary Materials). The EBSDFs obtained for Y123-orthoII and Bi$_2$Sr$_2$CaCu$_2$O$_{8+\delta}$ (Bi-2212) at various doping levels and temperatures are shown in the left column of Fig. 2. Because the optically determined EBSDFs have been suggested to have a magnetic origin\cite{carbotte:1999,schachinger:2000,hwang:2004,norman:2004,hwang:2006,hwang:2007} we assumed that they are associated with the antiferromagnetic spin fluctuations. In the right column of Fig. 2, the fluctuating spins are presented with the average frequencies of the EBSDFs above the background spin at two temperatures (lowest and ambient) for each sample. The average spin fluctuation ($Fluc_{spin}(t)$) is $Fluc_{spin} = 0.14 \cos(\langle\Omega_1\rangle t)$.Note that the amplitude of the fluctuation is arbitrary. For the optimally doped sample, the average frequency at the lowest temperature was slightly higher than that at room temperature, while the results for the underdoped and overdoped samples were the opposite (see Fig. 4(A)). The temperature-dependent average frequencies for all samples can be found in the Supplementary Materials. The coherence length can be obtained using a similar approach to that described previously for BCS superconductors. However, there are some differences between BCS superconductors and cuprates. In cuprates, the charge transport exhibits strong anisotropy; the $ab$ plane transport is dominant to that along the $c$ axis\cite{ono:2003}. Superconductivity typically occurs in the two-dimensional CuO$_2$ (or $ab$) plane. Additionally, cuprates have anisotropic $d$-wave superconducting gaps. The timescale of the retarded interaction in the spin fluctuation is reduced by a factor of 2 because the Cooper pair should be in a spin singlet state, as shown in the lower part of Fig. 3, where two possible singlet pair formations with a timescale of half a period are displayed. The binding force for Cooper pair formation is directly related to the magnetic retarded interaction caused by the spin fluctuations. The magnetic force is most likely associated with an attractive spin–spin (or spin-up and spin-down) interaction. Because of these crucial differences, the coherence length formula used for BCS superconductors should be modified. The first modification is associated with the two-dimensionality. An electron can be paired with any other electron on a circle with the electron at the center; thus, the coherence length can be reduced by a factor of $2\pi$, which is a full angle in two dimensions. The second modification is associated with the anisotropic $d$-wave superconducting gap, specifically, $\Delta(\theta) = \Delta_0 \cos(2\theta)$, where $\Delta_0$ is the maximum gap along the anti-nodal direction and the angle ($\theta$) is measured from the anti-nodal direction as shown in the upper part of Fig. 3. In general, the coherence length is inversely proportional to the superconducting gap\cite{tinkham:1975}. Therefore, the coherence length is dependent on $\theta$, that is, $\xi(\theta)$. In this case, we need to average the coherence length over the angle in [0, $\pi/4$], as $\langle \xi \rangle_{\theta} \equiv (4/\pi)\int_{0}^{\pi/4} \xi(\theta) d\theta$. The angle-dependent coherence length can be written as, $\xi(\theta) = v_F(1/2)[2\pi/\langle\Omega_1\rangle](1/2\pi)(1/\cos(2\theta))$. Here, 1/2 comes from the timescale reduction by a factor of 2. The average coherence length can be written as $\langle \xi \rangle_{\theta}= v_F/(2\langle\Omega_1\rangle) \langle1/\cos(2\theta)\rangle_{\theta}$. Here, $\langle1/\cos(2\theta)\rangle_{\theta} =(2/\pi) \ln[1/\cos(2\theta)+\tan(2\theta)]$, which shows a logarithmic singularity along the nodal direction (or at $\theta = \pi/4$). To estimate the average coherence length, we assumed that $\langle1/\cos(2\theta)\rangle_{\theta}\cong$ 3.33. In this case, we used $\pi/4 \cong 0.78$ rad. Although some uncertainties exist in determining the experimentally reliable value of $\langle1/\cos(2\theta)\rangle_{\theta}$ the assumption might be reasonable if the size of the maximum superconducting gap and the experimental accessibility to the lowest frequency cutoff are considered; if $\Delta_0 =$ 30 meV, $\Delta$(0.78 rad) = 0.32 meV (or 2.6 cm$^{-1}$). In general, optical spectra exclude data below approximately 50 cm$^{-1}$, which is well above the $\Delta$(0.78 rad).

\begin{figure}[!htbp]
  \vspace*{-0.7 cm}%
  \centerline{\includegraphics[width=3.7 in]{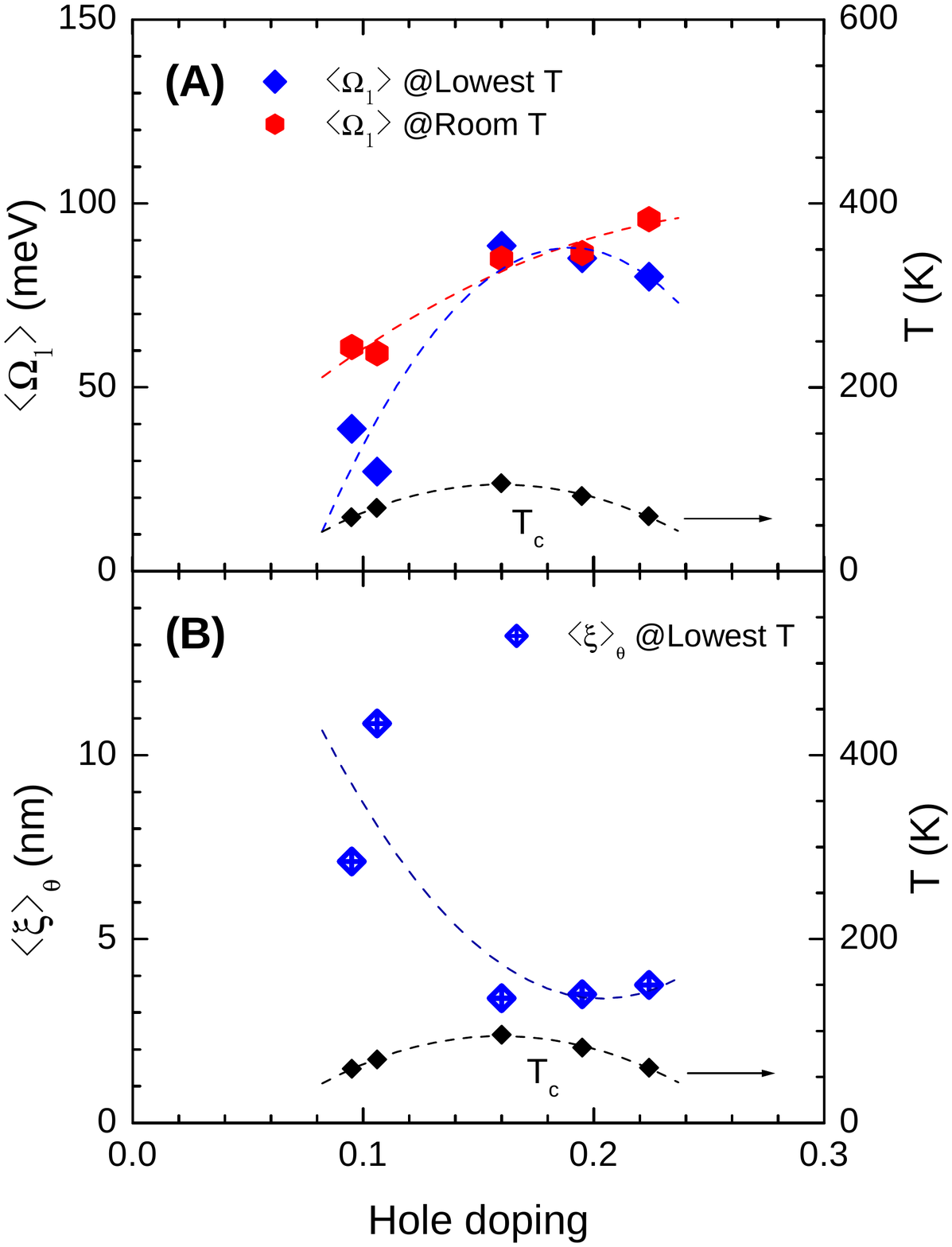}}%
  \vspace*{-1.2 cm}%
\caption{(Color online) Average frequencies of $I^2 B(\omega)$ and the coherence lengths for cuprates at various doping levels. (A) The doping-dependent average frequencies at two (lowest and room) temperatures. (B) The doping-dependent coherence lengths at the lowest temperature in the superconducting state.}
\label{fig4}
\end{figure}

The average coherence lengths of the cuprate samples were estimated at various doping levels and at two (lowest and room) temperatures using the modified formula for cuprates, as discussed previously: $\langle \xi \rangle_{\theta}= v_F/(2\langle\Omega_1\rangle)\times$ 3.33. The Fermi velocity of Bi-2212 is approximately 2.7$\times$10$^5$ m/s\cite{vishik:2010}. The same Fermi velocity value was used for all B-2212 samples. Meanwhile, the Fermi velocity of the Y123-orthoII sample is approximately 2.5$\times$ 10$^5$ m/s\cite{chiao:2000}. The average frequencies of $I^2B(\omega)$ at the two temperatures are shown in Fig. 4(A). As aforementioned, the average frequency at the lowest temperature was higher than that at room temperature only for the optimally doped sample. The corresponding estimated average coherence lengths at the lowest temperature are shown in Fig. 4(B). These results exhibits a strong doping dependence; the near-optimally doped coherence length is the shortest at approximately 3.5 nm. The estimated coherence lengths are reasonable compared with reported values; for example, $\xi_{ab} \simeq 1.6$ nm for an optimally doped YBa$_2$Cu$_3$O$_{6.90}$ with $T_c =$ 95 K and $\xi_{ab} =$ 3.8 nm for optimally doped La$_{1.85}$Sr$_{0.16}$CuO$_4$ with $T_c =$ 38 K\cite{mourachkine:2002}. In general, the coherence lengths of conventional BCS superconductors are hundreds of nanometers, whereas those of cuprates are in the nanometer range\cite{shen:2008}. It should be noted that these estimated coherence lengths might be overestimates because our lowest cutoff frequency ($\sim$50 cm$^{-1}$) is much higher than the one estimated ($\Delta$(0.78 rad)$\cong$ 2.6 cm$^{-1}$ for $\Delta_0 =$ 30 meV).

\begin{figure}[!htbp]
  \vspace*{-0.5 cm}%
  \centerline{\includegraphics[width=3.5 in]{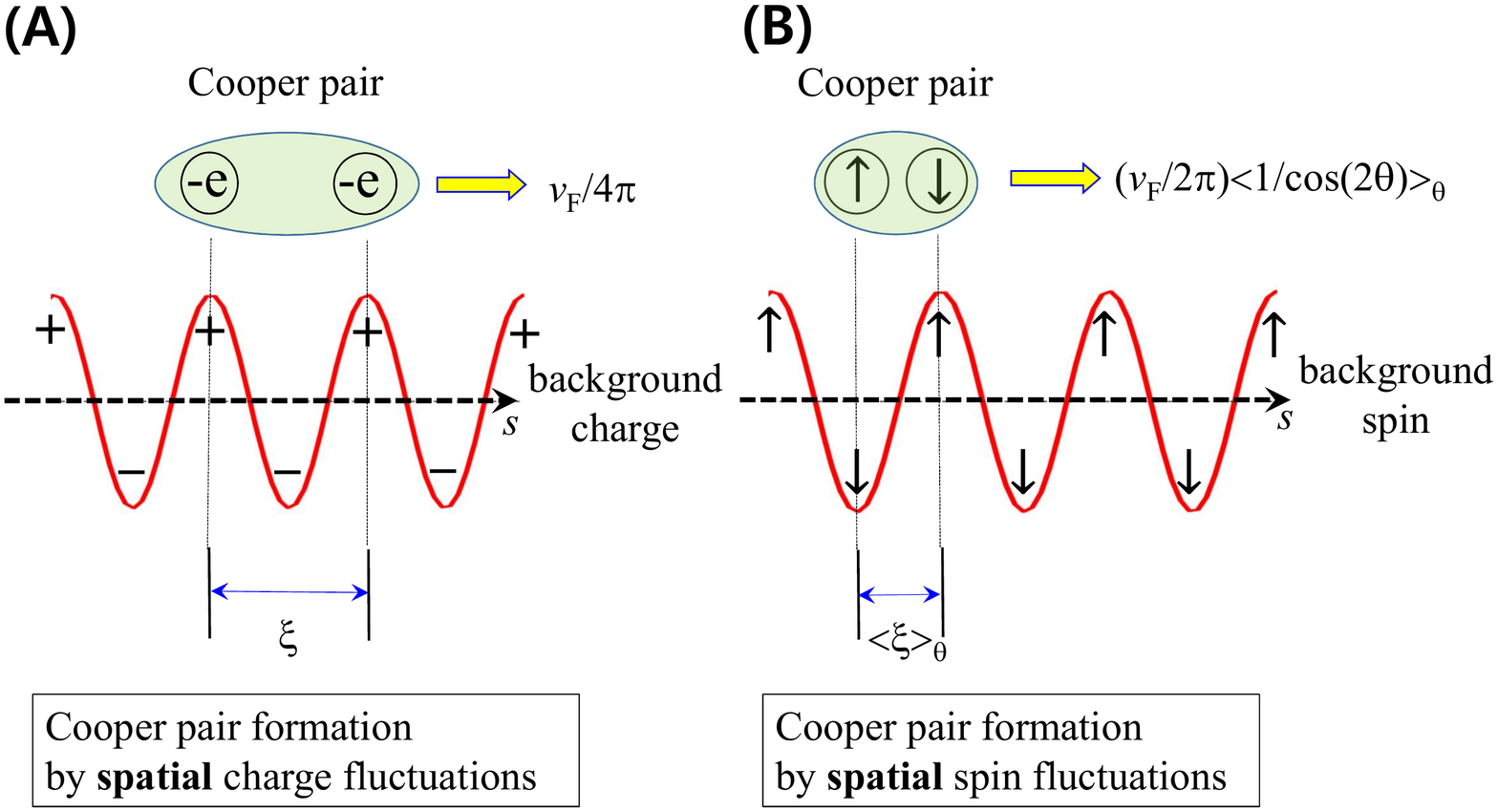}}%
  \vspace*{-1.0 cm}%
\caption{(Color online) {\it Spatial} charge/spin fluctuations and Cooper pair formations. (A) Cooper pair formation by electric retarded interaction caused by the charge fluctuations in the conventional BCS superconductors. (B) Cooper pair formation by magnetic retarded interaction caused by the spin fluctuations in the cuprate systems.}
\label{fig5}
\end{figure}

The Cooper pair formations are summarized through the spin (or charge) fluctuations in Fig. 5.The charge/spin fluctuations are the bonding glue for the formation of Cooper pairs moving with effective Fermi velocities, as shown in Figs. 5(A) and 5(B). The size of the Cooper pair (or coherence length) should adequately match the average period of {\it spatial} spin (or charge) fluctuations. The high-temperature superconductivity can be understood through the microscopic mechanism, in which spin-1 bosons with an antiferromagnetic spin texture are involved. The coherence lengths obtained from the experimental EBSDFs are consistent with those obtained from other experiments. The concept of the microscopic mechanism of Cooper pair formation in cuprates is similar to that in conventional BCS superconductors, except for some detailed differences. The first is in the force-mediating boson, which is associated with the spin fluctuations instead of the charge fluctuations in BCS superconductors. Other differences include the shorter timescale of spin fluctuations by one order and slower Fermi velocities by one order compared with the corresponding quantities for conventional superconductors. These differences result in shorter coherence lengths of cuprates by two orders compared with those of BCS superconductors.

One interesting question might be how the microscopic pairing information is encoded in the measured optical spectra. The optical spectra show the optical transitions averaged over the $k$-space, which are the transitions of electrons from occupied states below the Fermi level to unoccupied states above it. When the bands are renormalized by correlations between electrons caused by many-body effects, the band renormalization effects will be naturally encoded in the measured spectrum indirectly in the case of spin (or magnetic) fluctuations. From an appropriate analysis (or decoding) process, the correlation spectrum can be extracted from the measured spectrum, which is the EBSDF spectrum. The coherence length obtained from the timescale in the extracted EBSDF spectrum and the Fermi velocity is reasonable. In addition, the superconducting transition temperatures estimated from the EBSDF using a generalized McMillan formula are acceptable\cite{hwang:2008c,hwang:2011}. Therefore, the EBSDF may provide conclusive evidence for high-temperature superconductivity. In particular, using optical spectroscopy, the EBSDFs of almost all cuprates can be obtained because optical spectroscopy is sensitive to small signals, provides bulk properties, and covers a wide spectral range with a high energy resolution. Therefore, it can be a bridge between other experimental techniques, such as angle-resolved photoemission spectroscopy (surface sensitive) and inelastic neutron scattering (small scattering cross section). We expect that this approach can be applied to obtain the coherence lengths from EBSDFs established by other experimental techniques. We hope that our results will help researchers to conclusively establish the microscopic pairing mechanism for high-temperature superconductors, including cuprates and Fe-based superconductors.

\acknowledgments The author acknowledges Prof. Choi at Department of Physics, Sungkyunkwan University for his helpful comments on the manuscript. JH also acknowledges the financial support from the National Research Foundation of Korea (NRFK Grant No. 2019R1A6A1007307912).

\bibliographystyle{apsrev4-1}
\bibliography{bib}

\begin{thebibliography}{37}%
\makeatletter
\providecommand \@ifxundefined [1]{%
 \@ifx{#1\undefined}
}%
\providecommand \@ifnum [1]{%
 \ifnum #1\expandafter \@firstoftwo
 \else \expandafter \@secondoftwo
 \fi
}%
\providecommand \@ifx [1]{%
 \ifx #1\expandafter \@firstoftwo
 \else \expandafter \@secondoftwo
 \fi
}%
\providecommand \natexlab [1]{#1}%
\providecommand \enquote  [1]{``#1''}%
\providecommand \bibnamefont  [1]{#1}%
\providecommand \bibfnamefont [1]{#1}%
\providecommand \citenamefont [1]{#1}%
\providecommand \href@noop [0]{\@secondoftwo}%
\providecommand \href [0]{\begingroup \@sanitize@url \@href}%
\providecommand \@href[1]{\@@startlink{#1}\@@href}%
\providecommand \@@href[1]{\endgroup#1\@@endlink}%
\providecommand \@sanitize@url [0]{\catcode `\\12\catcode `\$12\catcode
  `\&12\catcode `\#12\catcode `\^12\catcode `\_12\catcode `\%12\relax}%
\providecommand \@@startlink[1]{}%
\providecommand \@@endlink[0]{}%
\providecommand \url  [0]{\begingroup\@sanitize@url \@url }%
\providecommand \@url [1]{\endgroup\@href {#1}{\urlprefix }}%
\providecommand \urlprefix  [0]{URL }%
\providecommand \Eprint [0]{\href }%
\providecommand \doibase [0]{http://dx.doi.org/}%
\providecommand \selectlanguage [0]{\@gobble}%
\providecommand \bibinfo  [0]{\@secondoftwo}%
\providecommand \bibfield  [0]{\@secondoftwo}%
\providecommand \translation [1]{[#1]}%
\providecommand \BibitemOpen [0]{}%
\providecommand \bibitemStop [0]{}%
\providecommand \bibitemNoStop [0]{.\EOS\space}%
\providecommand \EOS [0]{\spacefactor3000\relax}%
\providecommand \BibitemShut  [1]{\csname bibitem#1\endcsname}%
\let\auto@bib@innerbib\@empty
\bibitem [{\citenamefont {Bednorz}\ and\ \citenamefont
  {Muller}(1986)}]{bednorz:1986}%
  \BibitemOpen
  \bibfield  {author} {\bibinfo {author} {\bibfnamefont {T.~G.}\ \bibnamefont
  {Bednorz}}\ and\ \bibinfo {author} {\bibfnamefont {A.}~\bibnamefont
  {Muller}},\ }\href@noop {} {\bibfield  {journal} {\bibinfo  {journal} {Z.
  Phys. B}\ }\textbf {\bibinfo {volume} {64}},\ \bibinfo {pages} {189}
  (\bibinfo {year} {1986})}\BibitemShut {NoStop}%
\bibitem [{\citenamefont {Carbotte}\ \emph {et~al.}(1999)\citenamefont
  {Carbotte}, \citenamefont {Schachinger},\ and\ \citenamefont
  {Basov}}]{carbotte:1999}%
  \BibitemOpen
  \bibfield  {author} {\bibinfo {author} {\bibfnamefont {J.~P.}\ \bibnamefont
  {Carbotte}}, \bibinfo {author} {\bibfnamefont {E.}~\bibnamefont
  {Schachinger}}, \ and\ \bibinfo {author} {\bibfnamefont {D.~N.}\ \bibnamefont
  {Basov}},\ }\href@noop {} {\bibfield  {journal} {\bibinfo  {journal} {Nature
  (London)}\ }\textbf {\bibinfo {volume} {401}},\ \bibinfo {pages} {354}
  (\bibinfo {year} {1999})}\BibitemShut {NoStop}%
\bibitem [{\citenamefont {Schachinger}\ and\ \citenamefont
  {Carbotte}(2000)}]{schachinger:2000}%
  \BibitemOpen
  \bibfield  {author} {\bibinfo {author} {\bibfnamefont {E.}~\bibnamefont
  {Schachinger}}\ and\ \bibinfo {author} {\bibfnamefont {J.~P.}\ \bibnamefont
  {Carbotte}},\ }\href@noop {} {\bibfield  {journal} {\bibinfo  {journal}
  {Phys. Rev. B}\ }\textbf {\bibinfo {volume} {62}},\ \bibinfo {pages} {9054}
  (\bibinfo {year} {2000})}\BibitemShut {NoStop}%
\bibitem [{\citenamefont {Hwang}\ \emph {et~al.}(2006)\citenamefont {Hwang},
  \citenamefont {Yang}, \citenamefont {Timusk}, \citenamefont {Sharapov},
  \citenamefont {Carbotte}, \citenamefont {Bonn}, \citenamefont {Liang},\ and\
  \citenamefont {Hardy}}]{hwang:2006}%
  \BibitemOpen
  \bibfield  {author} {\bibinfo {author} {\bibfnamefont {J.}~\bibnamefont
  {Hwang}}, \bibinfo {author} {\bibfnamefont {J.}~\bibnamefont {Yang}},
  \bibinfo {author} {\bibfnamefont {T.}~\bibnamefont {Timusk}}, \bibinfo
  {author} {\bibfnamefont {S.~G.}\ \bibnamefont {Sharapov}}, \bibinfo {author}
  {\bibfnamefont {J.~P.}\ \bibnamefont {Carbotte}}, \bibinfo {author}
  {\bibfnamefont {D.~A.}\ \bibnamefont {Bonn}}, \bibinfo {author}
  {\bibfnamefont {R.}~\bibnamefont {Liang}}, \ and\ \bibinfo {author}
  {\bibfnamefont {W.~N.}\ \bibnamefont {Hardy}},\ }\href@noop {} {\bibfield
  {journal} {\bibinfo  {journal} {Phys. Rev. B}\ }\textbf {\bibinfo {volume}
  {73}},\ \bibinfo {pages} {014508} (\bibinfo {year} {2006})}\BibitemShut
  {NoStop}%
\bibitem [{\citenamefont {Zasadzinski}\ \emph {et~al.}(2006)\citenamefont
  {Zasadzinski}, \citenamefont {Ozyuzer}, \citenamefont {Coffey}, \citenamefont
  {Gray}, \citenamefont {Hinks},\ and\ \citenamefont
  {Kendziora}}]{zasadzinski:2006}%
  \BibitemOpen
  \bibfield  {author} {\bibinfo {author} {\bibfnamefont {J.~F.}\ \bibnamefont
  {Zasadzinski}}, \bibinfo {author} {\bibfnamefont {L.}~\bibnamefont
  {Ozyuzer}}, \bibinfo {author} {\bibfnamefont {L.}~\bibnamefont {Coffey}},
  \bibinfo {author} {\bibfnamefont {K.~E.}\ \bibnamefont {Gray}}, \bibinfo
  {author} {\bibfnamefont {D.~G.}\ \bibnamefont {Hinks}}, \ and\ \bibinfo
  {author} {\bibfnamefont {C.}~\bibnamefont {Kendziora}},\ }\href@noop {}
  {\bibfield  {journal} {\bibinfo  {journal} {Phys. Rev. Lett.}\ }\textbf
  {\bibinfo {volume} {96}},\ \bibinfo {pages} {017004} (\bibinfo {year}
  {2006})}\BibitemShut {NoStop}%
\bibitem [{\citenamefont {Lee}\ \emph {et~al.}(2006)\citenamefont {Lee},
  \citenamefont {Fujita}, \citenamefont {McElroy}, \citenamefont {Slezak},
  \citenamefont {Wang}, \citenamefont {Aiura}, \citenamefont {Bando},
  \citenamefont {Ishikado}, \citenamefont {Masui}, \citenamefont {Zhu},
  \citenamefont {Balatsky}, \citenamefont {Eisaki}, \citenamefont {Uchida},\
  and\ \citenamefont {Davis}}]{lee:2006}%
  \BibitemOpen
  \bibfield  {author} {\bibinfo {author} {\bibfnamefont {J.}~\bibnamefont
  {Lee}}, \bibinfo {author} {\bibfnamefont {K.}~\bibnamefont {Fujita}},
  \bibinfo {author} {\bibfnamefont {K.}~\bibnamefont {McElroy}}, \bibinfo
  {author} {\bibfnamefont {J.~A.}\ \bibnamefont {Slezak}}, \bibinfo {author}
  {\bibfnamefont {M.}~\bibnamefont {Wang}}, \bibinfo {author} {\bibfnamefont
  {Y.}~\bibnamefont {Aiura}}, \bibinfo {author} {\bibfnamefont
  {H.}~\bibnamefont {Bando}}, \bibinfo {author} {\bibfnamefont
  {M.}~\bibnamefont {Ishikado}}, \bibinfo {author} {\bibfnamefont
  {T.}~\bibnamefont {Masui}}, \bibinfo {author} {\bibfnamefont {J.-X.}\
  \bibnamefont {Zhu}}, \bibinfo {author} {\bibfnamefont {A.~V.}\ \bibnamefont
  {Balatsky}}, \bibinfo {author} {\bibfnamefont {H.}~\bibnamefont {Eisaki}},
  \bibinfo {author} {\bibfnamefont {S.}~\bibnamefont {Uchida}}, \ and\ \bibinfo
  {author} {\bibfnamefont {J.~C.}\ \bibnamefont {Davis}},\ }\href@noop {}
  {\bibfield  {journal} {\bibinfo  {journal} {Nature (London)}\ }\textbf
  {\bibinfo {volume} {442}},\ \bibinfo {pages} {546} (\bibinfo {year}
  {2006})}\BibitemShut {NoStop}%
\bibitem [{\citenamefont {Valla}\ \emph {et~al.}(2007)\citenamefont {Valla},
  \citenamefont {Kidd}, \citenamefont {Yin}, \citenamefont {Gu}, \citenamefont
  {Johnson}, \citenamefont {Pan},\ and\ \citenamefont {Fedorov}}]{valla:2007}%
  \BibitemOpen
  \bibfield  {author} {\bibinfo {author} {\bibfnamefont {T.}~\bibnamefont
  {Valla}}, \bibinfo {author} {\bibfnamefont {T.~E.}\ \bibnamefont {Kidd}},
  \bibinfo {author} {\bibfnamefont {W.-G.}\ \bibnamefont {Yin}}, \bibinfo
  {author} {\bibfnamefont {G.~D.}\ \bibnamefont {Gu}}, \bibinfo {author}
  {\bibfnamefont {P.~D.}\ \bibnamefont {Johnson}}, \bibinfo {author}
  {\bibfnamefont {Z.-H.}\ \bibnamefont {Pan}}, \ and\ \bibinfo {author}
  {\bibfnamefont {A.~V.}\ \bibnamefont {Fedorov}},\ }\href@noop {} {\bibfield
  {journal} {\bibinfo  {journal} {Phys. Rev. Lett.}\ }\textbf {\bibinfo
  {volume} {98}},\ \bibinfo {pages} {167003} (\bibinfo {year}
  {2007})}\BibitemShut {NoStop}%
\bibitem [{\citenamefont {Hwang}\ \emph
  {et~al.}(2007{\natexlab{a}})\citenamefont {Hwang}, \citenamefont {Timusk},
  \citenamefont {Schachinger},\ and\ \citenamefont {Carbotte}}]{hwang:2007}%
  \BibitemOpen
  \bibfield  {author} {\bibinfo {author} {\bibfnamefont {J.}~\bibnamefont
  {Hwang}}, \bibinfo {author} {\bibfnamefont {T.}~\bibnamefont {Timusk}},
  \bibinfo {author} {\bibfnamefont {E.}~\bibnamefont {Schachinger}}, \ and\
  \bibinfo {author} {\bibfnamefont {J.~P.}\ \bibnamefont {Carbotte}},\
  }\href@noop {} {\bibfield  {journal} {\bibinfo  {journal} {Phys. Rev. B}\
  }\textbf {\bibinfo {volume} {75}},\ \bibinfo {pages} {144508} (\bibinfo
  {year} {2007}{\natexlab{a}})}\BibitemShut {NoStop}%
\bibitem [{\citenamefont {Hwang}\ \emph
  {et~al.}(2007{\natexlab{b}})\citenamefont {Hwang}, \citenamefont {Timusk},\
  and\ \citenamefont {Carbotte}}]{hwang:2007ab}%
  \BibitemOpen
  \bibfield  {author} {\bibinfo {author} {\bibfnamefont {J.}~\bibnamefont
  {Hwang}}, \bibinfo {author} {\bibfnamefont {T.}~\bibnamefont {Timusk}}, \
  and\ \bibinfo {author} {\bibfnamefont {J.~P.}\ \bibnamefont {Carbotte}},\
  }\href@noop {} {\bibfield  {journal} {\bibinfo  {journal} {Nature (London)}\
  }\textbf {\bibinfo {volume} {446}},\ \bibinfo {pages} {E3} (\bibinfo {year}
  {2007}{\natexlab{b}})}\BibitemShut {NoStop}%
\bibitem [{\citenamefont {van Heumen}\ \emph
  {et~al.}(2009{\natexlab{a}})\citenamefont {van Heumen}, \citenamefont
  {Muhlethaler}, \citenamefont {Kuzmenko}, \citenamefont {Eisaki},
  \citenamefont {Meevasana}, \citenamefont {Greven},\ and\ \citenamefont
  {van~der Marel}}]{heumen:2009}%
  \BibitemOpen
  \bibfield  {author} {\bibinfo {author} {\bibfnamefont {E.}~\bibnamefont {van
  Heumen}}, \bibinfo {author} {\bibfnamefont {E.}~\bibnamefont {Muhlethaler}},
  \bibinfo {author} {\bibfnamefont {A.~B.}\ \bibnamefont {Kuzmenko}}, \bibinfo
  {author} {\bibfnamefont {H.}~\bibnamefont {Eisaki}}, \bibinfo {author}
  {\bibfnamefont {W.}~\bibnamefont {Meevasana}}, \bibinfo {author}
  {\bibfnamefont {M.}~\bibnamefont {Greven}}, \ and\ \bibinfo {author}
  {\bibfnamefont {D.}~\bibnamefont {van~der Marel}},\ }\href@noop {} {\bibfield
   {journal} {\bibinfo  {journal} {Phys. Rev. B}\ }\textbf {\bibinfo {volume}
  {79}},\ \bibinfo {pages} {184512} (\bibinfo {year}
  {2009}{\natexlab{a}})}\BibitemShut {NoStop}%
\bibitem [{\citenamefont {Hwang}(2011)}]{hwang:2011}%
  \BibitemOpen
  \bibfield  {author} {\bibinfo {author} {\bibfnamefont {J.}~\bibnamefont
  {Hwang}},\ }\href@noop {} {\bibfield  {journal} {\bibinfo  {journal} {Phys.
  Rev. B}\ }\textbf {\bibinfo {volume} {83}},\ \bibinfo {pages} {014507}
  (\bibinfo {year} {2011})}\BibitemShut {NoStop}%
\bibitem [{\citenamefont {Carbotte}\ \emph {et~al.}(2011)\citenamefont
  {Carbotte}, \citenamefont {Timusk},\ and\ \citenamefont
  {Hwang}}]{carbotte:2011}%
  \BibitemOpen
  \bibfield  {author} {\bibinfo {author} {\bibfnamefont {J.~P.}\ \bibnamefont
  {Carbotte}}, \bibinfo {author} {\bibfnamefont {T.}~\bibnamefont {Timusk}}, \
  and\ \bibinfo {author} {\bibfnamefont {J.}~\bibnamefont {Hwang}},\
  }\href@noop {} {\bibfield  {journal} {\bibinfo  {journal} {Reports on
  Progress in Physics}\ }\textbf {\bibinfo {volume} {74}},\ \bibinfo {pages}
  {066501} (\bibinfo {year} {2011})}\BibitemShut {NoStop}%
\bibitem [{\citenamefont {Ahmadi}\ \emph {et~al.}(2011)\citenamefont {Ahmadi},
  \citenamefont {Coffey}, \citenamefont {Zasadzinski}, \citenamefont
  {Miyakawa},\ and\ \citenamefont {Ozyuzer}}]{zasadzinski:2011}%
  \BibitemOpen
  \bibfield  {author} {\bibinfo {author} {\bibfnamefont {O.}~\bibnamefont
  {Ahmadi}}, \bibinfo {author} {\bibfnamefont {L.}~\bibnamefont {Coffey}},
  \bibinfo {author} {\bibfnamefont {J.~F.}\ \bibnamefont {Zasadzinski}},
  \bibinfo {author} {\bibfnamefont {N.}~\bibnamefont {Miyakawa}}, \ and\
  \bibinfo {author} {\bibfnamefont {L.}~\bibnamefont {Ozyuzer}},\ }\href@noop
  {} {\bibfield  {journal} {\bibinfo  {journal} {Phys. Rev. Lett.}\ }\textbf
  {\bibinfo {volume} {106}},\ \bibinfo {pages} {167005} (\bibinfo {year}
  {2011})}\BibitemShut {NoStop}%
\bibitem [{\citenamefont {Johnson}\ \emph {et~al.}(2001)\citenamefont
  {Johnson}, \citenamefont {Valla}, \citenamefont {Fedorov}, \citenamefont
  {Yusof}, \citenamefont {Wells}, \citenamefont {Li}, \citenamefont
  {Moodenbaugh}, \citenamefont {Gu}, \citenamefont {Koshizuka}, \citenamefont
  {Kendziora}, \citenamefont {Jian},\ and\ \citenamefont
  {Hinks}}]{johnson:2001}%
  \BibitemOpen
  \bibfield  {author} {\bibinfo {author} {\bibfnamefont {P.~D.}\ \bibnamefont
  {Johnson}}, \bibinfo {author} {\bibfnamefont {T.}~\bibnamefont {Valla}},
  \bibinfo {author} {\bibfnamefont {A.~V.}\ \bibnamefont {Fedorov}}, \bibinfo
  {author} {\bibfnamefont {Z.}~\bibnamefont {Yusof}}, \bibinfo {author}
  {\bibfnamefont {B.~O.}\ \bibnamefont {Wells}}, \bibinfo {author}
  {\bibfnamefont {Q.}~\bibnamefont {Li}}, \bibinfo {author} {\bibfnamefont
  {A.~R.}\ \bibnamefont {Moodenbaugh}}, \bibinfo {author} {\bibfnamefont
  {G.~D.}\ \bibnamefont {Gu}}, \bibinfo {author} {\bibfnamefont
  {N.}~\bibnamefont {Koshizuka}}, \bibinfo {author} {\bibfnamefont
  {C.}~\bibnamefont {Kendziora}}, \bibinfo {author} {\bibfnamefont
  {S.}~\bibnamefont {Jian}}, \ and\ \bibinfo {author} {\bibfnamefont {D.~G.}\
  \bibnamefont {Hinks}},\ }\href@noop {} {\bibfield  {journal} {\bibinfo
  {journal} {Phys. Rev. Lett.}\ }\textbf {\bibinfo {volume} {87}},\ \bibinfo
  {pages} {177007} (\bibinfo {year} {2001})}\BibitemShut {NoStop}%
\bibitem [{\citenamefont {Zasadzinski}\ \emph {et~al.}(2001)\citenamefont
  {Zasadzinski}, \citenamefont {Ozyuzer}, \citenamefont {Miyakawa},
  \citenamefont {Gray}, \citenamefont {Hinks},\ and\ \citenamefont
  {Kendziora}}]{zasadzinski:2001}%
  \BibitemOpen
  \bibfield  {author} {\bibinfo {author} {\bibfnamefont {J.~F.}\ \bibnamefont
  {Zasadzinski}}, \bibinfo {author} {\bibfnamefont {L.}~\bibnamefont
  {Ozyuzer}}, \bibinfo {author} {\bibfnamefont {N.}~\bibnamefont {Miyakawa}},
  \bibinfo {author} {\bibfnamefont {K.~E.}\ \bibnamefont {Gray}}, \bibinfo
  {author} {\bibfnamefont {D.~G.}\ \bibnamefont {Hinks}}, \ and\ \bibinfo
  {author} {\bibfnamefont {C.}~\bibnamefont {Kendziora}},\ }\href@noop {}
  {\bibfield  {journal} {\bibinfo  {journal} {Phys. Rev. Lett.}\ }\textbf
  {\bibinfo {volume} {87}},\ \bibinfo {pages} {067005} (\bibinfo {year}
  {2001})}\BibitemShut {NoStop}%
\bibitem [{\citenamefont {Hwang}\ \emph {et~al.}(2004)\citenamefont {Hwang},
  \citenamefont {Timusk},\ and\ \citenamefont {Gu}}]{hwang:2004}%
  \BibitemOpen
  \bibfield  {author} {\bibinfo {author} {\bibfnamefont {J.}~\bibnamefont
  {Hwang}}, \bibinfo {author} {\bibfnamefont {T.}~\bibnamefont {Timusk}}, \
  and\ \bibinfo {author} {\bibfnamefont {G.~D.}\ \bibnamefont {Gu}},\
  }\href@noop {} {\bibfield  {journal} {\bibinfo  {journal} {Nature (London)}\
  }\textbf {\bibinfo {volume} {427}},\ \bibinfo {pages} {714} (\bibinfo {year}
  {2004})}\BibitemShut {NoStop}%
\bibitem [{\citenamefont {Norman}(2004)}]{norman:2004}%
  \BibitemOpen
  \bibfield  {author} {\bibinfo {author} {\bibfnamefont {M.}~\bibnamefont
  {Norman}},\ }\href@noop {} {\bibfield  {journal} {\bibinfo  {journal} {Nature
  (London)}\ }\textbf {\bibinfo {volume} {427}},\ \bibinfo {pages} {692}
  (\bibinfo {year} {2004})}\BibitemShut {NoStop}%
\bibitem [{\citenamefont {Dahm}\ \emph {et~al.}(2009)\citenamefont {Dahm},
  \citenamefont {Hinkov}, \citenamefont {Borisenko}, \citenamefont {Kordyuk},
  \citenamefont {Zabolotnyy}, \citenamefont {Fink}, \citenamefont {Buchner},
  \citenamefont {Scalapino}, \citenamefont {Hanke},\ and\ \citenamefont
  {Keimer}}]{dahm:2009}%
  \BibitemOpen
  \bibfield  {author} {\bibinfo {author} {\bibfnamefont {T.}~\bibnamefont
  {Dahm}}, \bibinfo {author} {\bibfnamefont {V.}~\bibnamefont {Hinkov}},
  \bibinfo {author} {\bibfnamefont {S.~V.}\ \bibnamefont {Borisenko}}, \bibinfo
  {author} {\bibfnamefont {A.~A.}\ \bibnamefont {Kordyuk}}, \bibinfo {author}
  {\bibfnamefont {V.~B.}\ \bibnamefont {Zabolotnyy}}, \bibinfo {author}
  {\bibfnamefont {J.}~\bibnamefont {Fink}}, \bibinfo {author} {\bibfnamefont
  {B.}~\bibnamefont {Buchner}}, \bibinfo {author} {\bibfnamefont {D.~J.}\
  \bibnamefont {Scalapino}}, \bibinfo {author} {\bibfnamefont {W.}~\bibnamefont
  {Hanke}}, \ and\ \bibinfo {author} {\bibfnamefont {B.}~\bibnamefont
  {Keimer}},\ }\href@noop {} {\bibfield  {journal} {\bibinfo  {journal} {Nat.
  Phys.}\ }\textbf {\bibinfo {volume} {5}},\ \bibinfo {pages} {217} (\bibinfo
  {year} {2009})}\BibitemShut {NoStop}%
\bibitem [{\citenamefont {van Heumen}\ \emph
  {et~al.}(2009{\natexlab{b}})\citenamefont {van Heumen}, \citenamefont
  {Meevasana}, \citenamefont {Kuzmenko}, \citenamefont {Eisaki},\ and\
  \citenamefont {van derMarel}}]{heumen:2009a}%
  \BibitemOpen
  \bibfield  {author} {\bibinfo {author} {\bibfnamefont {E.}~\bibnamefont {van
  Heumen}}, \bibinfo {author} {\bibfnamefont {W.}~\bibnamefont {Meevasana}},
  \bibinfo {author} {\bibfnamefont {A.~B.}\ \bibnamefont {Kuzmenko}}, \bibinfo
  {author} {\bibfnamefont {H.}~\bibnamefont {Eisaki}}, \ and\ \bibinfo {author}
  {\bibfnamefont {D.}~\bibnamefont {van derMarel}},\ }\href@noop {} {\bibfield
  {journal} {\bibinfo  {journal} {New Journal of Physics}\ }\textbf {\bibinfo
  {volume} {11}},\ \bibinfo {pages} {055067} (\bibinfo {year}
  {2009}{\natexlab{b}})}\BibitemShut {NoStop}%
\bibitem [{\citenamefont {Hwang}(2015)}]{hwang:2015a}%
  \BibitemOpen
  \bibfield  {author} {\bibinfo {author} {\bibfnamefont {J.}~\bibnamefont
  {Hwang}},\ }\href@noop {} {\bibfield  {journal} {\bibinfo  {journal} {J.
  Phys.: Condens. Matter}\ }\textbf {\bibinfo {volume} {27}},\ \bibinfo {pages}
  {085701} (\bibinfo {year} {2015})}\BibitemShut {NoStop}%
\bibitem [{\citenamefont {Hwang}\ \emph
  {et~al.}(2008{\natexlab{a}})\citenamefont {Hwang}, \citenamefont
  {Schachinger}, \citenamefont {Carbotte}, \citenamefont {Gao}, \citenamefont
  {Tanner},\ and\ \citenamefont {Timusk}}]{hwang:2008c}%
  \BibitemOpen
  \bibfield  {author} {\bibinfo {author} {\bibfnamefont {J.}~\bibnamefont
  {Hwang}}, \bibinfo {author} {\bibfnamefont {E.}~\bibnamefont {Schachinger}},
  \bibinfo {author} {\bibfnamefont {J.~P.}\ \bibnamefont {Carbotte}}, \bibinfo
  {author} {\bibfnamefont {F.}~\bibnamefont {Gao}}, \bibinfo {author}
  {\bibfnamefont {D.~B.}\ \bibnamefont {Tanner}}, \ and\ \bibinfo {author}
  {\bibfnamefont {T.}~\bibnamefont {Timusk}},\ }\href@noop {} {\bibfield
  {journal} {\bibinfo  {journal} {Phys. Rev. Lett.}\ }\textbf {\bibinfo
  {volume} {100}},\ \bibinfo {pages} {137005} (\bibinfo {year}
  {2008}{\natexlab{a}})}\BibitemShut {NoStop}%
\bibitem [{\citenamefont {McMillan}\ and\ \citenamefont
  {Rowell}(1965)}]{mcmillan:1965}%
  \BibitemOpen
  \bibfield  {author} {\bibinfo {author} {\bibfnamefont {W.~L.}\ \bibnamefont
  {McMillan}}\ and\ \bibinfo {author} {\bibfnamefont {J.~M.}\ \bibnamefont
  {Rowell}},\ }\href@noop {} {\bibfield  {journal} {\bibinfo  {journal} {Phys.
  Rev. Lett.}\ }\textbf {\bibinfo {volume} {14}},\ \bibinfo {pages} {108}
  (\bibinfo {year} {1965})}\BibitemShut {NoStop}%
\bibitem [{\citenamefont {Farnworth}\ and\ \citenamefont
  {Timusk}(1974)}]{farnworth:1974}%
  \BibitemOpen
  \bibfield  {author} {\bibinfo {author} {\bibfnamefont {B.}~\bibnamefont
  {Farnworth}}\ and\ \bibinfo {author} {\bibfnamefont {T.}~\bibnamefont
  {Timusk}},\ }\href@noop {} {\bibfield  {journal} {\bibinfo  {journal} {Phys.
  Rev. B}\ }\textbf {\bibinfo {volume} {10}},\ \bibinfo {pages} {2799}
  (\bibinfo {year} {1974})}\BibitemShut {NoStop}%
\bibitem [{\citenamefont {Farnworth}\ and\ \citenamefont
  {Timusk}(1976)}]{farnworth:1976}%
  \BibitemOpen
  \bibfield  {author} {\bibinfo {author} {\bibfnamefont {B.}~\bibnamefont
  {Farnworth}}\ and\ \bibinfo {author} {\bibfnamefont {T.}~\bibnamefont
  {Timusk}},\ }\href@noop {} {\bibfield  {journal} {\bibinfo  {journal} {Phys.
  Rev. B}\ }\textbf {\bibinfo {volume} {14}},\ \bibinfo {pages} {5119}
  (\bibinfo {year} {1976})}\BibitemShut {NoStop}%
\bibitem [{\citenamefont {Tomlinson}\ and\ \citenamefont
  {Carbotte}(1976)}]{tomlinson:1976}%
  \BibitemOpen
  \bibfield  {author} {\bibinfo {author} {\bibfnamefont {P.~G.}\ \bibnamefont
  {Tomlinson}}\ and\ \bibinfo {author} {\bibfnamefont {J.~P.}\ \bibnamefont
  {Carbotte}},\ }\href@noop {} {\bibfield  {journal} {\bibinfo  {journal}
  {Phys. Rev. B}\ }\textbf {\bibinfo {volume} {13}},\ \bibinfo {pages} {4738}
  (\bibinfo {year} {1976})}\BibitemShut {NoStop}%
\bibitem [{\citenamefont {Lykken}\ \emph {et~al.}(1971)\citenamefont {Lykken},
  \citenamefont {Geiger}, \citenamefont {Dy},\ and\ \citenamefont
  {Mitchell}}]{lykken:1971}%
  \BibitemOpen
  \bibfield  {author} {\bibinfo {author} {\bibfnamefont {G.~I.}\ \bibnamefont
  {Lykken}}, \bibinfo {author} {\bibfnamefont {A.~L.}\ \bibnamefont {Geiger}},
  \bibinfo {author} {\bibfnamefont {K.~S.}\ \bibnamefont {Dy}}, \ and\ \bibinfo
  {author} {\bibfnamefont {E.~N.}\ \bibnamefont {Mitchell}},\ }\href@noop {}
  {\bibfield  {journal} {\bibinfo  {journal} {Phys. Rev. B}\ }\textbf {\bibinfo
  {volume} {4}},\ \bibinfo {pages} {1523} (\bibinfo {year} {1971})}\BibitemShut
  {NoStop}%
\bibitem [{\citenamefont {Ashcroft}\ and\ \citenamefont
  {Mermin}(1976)}]{ashcroft}%
  \BibitemOpen
  \bibfield  {author} {\bibinfo {author} {\bibfnamefont {N.~W.}\ \bibnamefont
  {Ashcroft}}\ and\ \bibinfo {author} {\bibfnamefont {N.~D.}\ \bibnamefont
  {Mermin}},\ }\href@noop {} {\emph {\bibinfo {title} {Solid State Physics}}}\
  (\bibinfo  {publisher} {Saunders College Publishing},\ \bibinfo {year}
  {1976})\BibitemShut {NoStop}%
\bibitem [{\citenamefont {Gasparovic}\ and\ \citenamefont
  {McLean}(1970)}]{gasparovic:1970}%
  \BibitemOpen
  \bibfield  {author} {\bibinfo {author} {\bibfnamefont {R.~F.}\ \bibnamefont
  {Gasparovic}}\ and\ \bibinfo {author} {\bibfnamefont {W.~L.}\ \bibnamefont
  {McLean}},\ }\href@noop {} {\bibfield  {journal} {\bibinfo  {journal} {Phys.
  Rev. B}\ }\textbf {\bibinfo {volume} {2}},\ \bibinfo {pages} {2519} (\bibinfo
  {year} {1970})}\BibitemShut {NoStop}%
\bibitem [{\citenamefont {Hwang}\ \emph
  {et~al.}(2008{\natexlab{b}})\citenamefont {Hwang}, \citenamefont {Carbotte},\
  and\ \citenamefont {Timusk}}]{hwang:2008}%
  \BibitemOpen
  \bibfield  {author} {\bibinfo {author} {\bibfnamefont {J.}~\bibnamefont
  {Hwang}}, \bibinfo {author} {\bibfnamefont {J.~P.}\ \bibnamefont {Carbotte}},
  \ and\ \bibinfo {author} {\bibfnamefont {T.}~\bibnamefont {Timusk}},\
  }\href@noop {} {\bibfield  {journal} {\bibinfo  {journal} {Phys. Rev. Lett.}\
  }\textbf {\bibinfo {volume} {100}},\ \bibinfo {pages} {177005} (\bibinfo
  {year} {2008}{\natexlab{b}})}\BibitemShut {NoStop}%
\bibitem [{\citenamefont {Hwang}(2016)}]{hwang:2016a}%
  \BibitemOpen
  \bibfield  {author} {\bibinfo {author} {\bibfnamefont {J.}~\bibnamefont
  {Hwang}},\ }\href@noop {} {\bibfield  {journal} {\bibinfo  {journal}
  {Scientific Reports}\ }\textbf {\bibinfo {volume} {6}},\ \bibinfo {pages}
  {23647} (\bibinfo {year} {2016})}\BibitemShut {NoStop}%
\bibitem [{\citenamefont {Schachinger}\ \emph {et~al.}(2006)\citenamefont
  {Schachinger}, \citenamefont {Neuber},\ and\ \citenamefont
  {Carbotte}}]{schachinger:2006}%
  \BibitemOpen
  \bibfield  {author} {\bibinfo {author} {\bibfnamefont {E.}~\bibnamefont
  {Schachinger}}, \bibinfo {author} {\bibfnamefont {D.}~\bibnamefont {Neuber}},
  \ and\ \bibinfo {author} {\bibfnamefont {J.~P.}\ \bibnamefont {Carbotte}},\
  }\href@noop {} {\bibfield  {journal} {\bibinfo  {journal} {Phys. Rev. B}\
  }\textbf {\bibinfo {volume} {73}},\ \bibinfo {pages} {184507} (\bibinfo
  {year} {2006})}\BibitemShut {NoStop}%
\bibitem [{\citenamefont {Ono}\ and\ \citenamefont {Ando}(2003)}]{ono:2003}%
  \BibitemOpen
  \bibfield  {author} {\bibinfo {author} {\bibfnamefont {S.}~\bibnamefont
  {Ono}}\ and\ \bibinfo {author} {\bibfnamefont {Y.}~\bibnamefont {Ando}},\
  }\href@noop {} {\bibfield  {journal} {\bibinfo  {journal} {Phys. Rev. B}\
  }\textbf {\bibinfo {volume} {67}},\ \bibinfo {pages} {104512} (\bibinfo
  {year} {2003})}\BibitemShut {NoStop}%
\bibitem [{\citenamefont {Tinkham}(1975)}]{tinkham:1975}%
  \BibitemOpen
  \bibfield  {author} {\bibinfo {author} {\bibfnamefont {M.}~\bibnamefont
  {Tinkham}},\ }\href@noop {} {\emph {\bibinfo {title} {Introduction to
  Superconductivity}}}\ (\bibinfo  {publisher} {McGraw-Hill Book Co., New
  York},\ \bibinfo {year} {1975})\BibitemShut {NoStop}%
\bibitem [{\citenamefont {Vishik}\ \emph {et~al.}(2010)\citenamefont {Vishik},
  \citenamefont {Lee}, \citenamefont {Schmitt}, \citenamefont {Moritz},
  \citenamefont {Sasagawa}, \citenamefont {Uchida}, \citenamefont {Fujita},
  \citenamefont {Ishida}, \citenamefont {Zhang}, \citenamefont {Devereaux},\
  and\ \citenamefont {Shen}}]{vishik:2010}%
  \BibitemOpen
  \bibfield  {author} {\bibinfo {author} {\bibfnamefont {I.~M.}\ \bibnamefont
  {Vishik}}, \bibinfo {author} {\bibfnamefont {W.~S.}\ \bibnamefont {Lee}},
  \bibinfo {author} {\bibfnamefont {F.}~\bibnamefont {Schmitt}}, \bibinfo
  {author} {\bibfnamefont {B.}~\bibnamefont {Moritz}}, \bibinfo {author}
  {\bibfnamefont {T.}~\bibnamefont {Sasagawa}}, \bibinfo {author}
  {\bibfnamefont {S.}~\bibnamefont {Uchida}}, \bibinfo {author} {\bibfnamefont
  {K.}~\bibnamefont {Fujita}}, \bibinfo {author} {\bibfnamefont
  {S.}~\bibnamefont {Ishida}}, \bibinfo {author} {\bibfnamefont
  {C.}~\bibnamefont {Zhang}}, \bibinfo {author} {\bibfnamefont {T.~P.}\
  \bibnamefont {Devereaux}}, \ and\ \bibinfo {author} {\bibfnamefont {Z.~X.}\
  \bibnamefont {Shen}},\ }\href@noop {} {\bibfield  {journal} {\bibinfo
  {journal} {Phys. Rev. Lett.}\ }\textbf {\bibinfo {volume} {104}},\ \bibinfo
  {pages} {207002} (\bibinfo {year} {2010})}\BibitemShut {NoStop}%
\bibitem [{\citenamefont {Chiao}\ \emph {et~al.}(2000)\citenamefont {Chiao},
  \citenamefont {Hill}, \citenamefont {Lupien}, \citenamefont {Taillefer},
  \citenamefont {Lambert}, \citenamefont {Gagnon},\ and\ \citenamefont
  {Fournier}}]{chiao:2000}%
  \BibitemOpen
  \bibfield  {author} {\bibinfo {author} {\bibfnamefont {M.}~\bibnamefont
  {Chiao}}, \bibinfo {author} {\bibfnamefont {R.~W.}\ \bibnamefont {Hill}},
  \bibinfo {author} {\bibfnamefont {C.}~\bibnamefont {Lupien}}, \bibinfo
  {author} {\bibfnamefont {L.}~\bibnamefont {Taillefer}}, \bibinfo {author}
  {\bibfnamefont {P.}~\bibnamefont {Lambert}}, \bibinfo {author} {\bibfnamefont
  {R.}~\bibnamefont {Gagnon}}, \ and\ \bibinfo {author} {\bibfnamefont
  {P.}~\bibnamefont {Fournier}},\ }\href@noop {} {\bibfield  {journal}
  {\bibinfo  {journal} {Phys. Rev. B}\ }\textbf {\bibinfo {volume} {62}},\
  \bibinfo {pages} {3554} (\bibinfo {year} {2000})}\BibitemShut {NoStop}%
\bibitem [{\citenamefont {Mourachkine}(2002)}]{mourachkine:2002}%
  \BibitemOpen
  \bibfield  {author} {\bibinfo {author} {\bibfnamefont {A.}~\bibnamefont
  {Mourachkine}},\ }\href@noop {} {\emph {\bibinfo {title} {High-Temperature
  Superconductivity in cuprates: The Nonlinear Mechanism and Tunneling
  Measurements}}}\ (\bibinfo  {publisher} {Kluwer Academic Publishers},\
  \bibinfo {year} {2002})\ \bibinfo {note} {(Note: Key material in Table 3.2 on
  page 60)}\BibitemShut {NoStop}%
\bibitem [{\citenamefont {Shen}\ and\ \citenamefont {Davis}(2008)}]{shen:2008}%
  \BibitemOpen
  \bibfield  {author} {\bibinfo {author} {\bibfnamefont {K.~M.}\ \bibnamefont
  {Shen}}\ and\ \bibinfo {author} {\bibfnamefont {J.~C.~S.}\ \bibnamefont
  {Davis}},\ }\href@noop {} {\bibfield  {journal} {\bibinfo  {journal}
  {Materials Today}\ }\textbf {\bibinfo {volume} {11}},\ \bibinfo {pages} {14}
  (\bibinfo {year} {2008})}\BibitemShut {NoStop}%
\end{thebibliography}%

\end{document}